\documentclass[11pt]{article}
\usepackage[totalwidth=460truept,totalheight=600truept]{geometry}
\usepackage{latexsym,amssymb,amsmath,graphicx}
\usepackage[T1]{fontenc}
\usepackage{color}

\global\arraycolsep=1truept
\newcommand{\beq}{\begin{equation}}
\newcommand{\eeq}{\end{equation}}
\def\bea{\begin{eqnarray}}
\def\eea{\end{eqnarray}}
\definecolor{darkred}{rgb}{.8,0,0}

\definecolor{darkblu}{rgb}{0,0,.8}

\begin{document}

%\hfill UB-ECM-PF-09/24

%\hfill UB-ICC-09/xx

\vskip 1.4truecm

\begin{center}
{\huge \textbf{ On the equivalence of the Einstein-Hilbert and the Einstein-Palatini formulations of general relativity for an arbitrary connection}}

\vskip 1.5truecm

\textsl{Naresh Dadhich}

\textsl{IUCAA, Post Bag 4, Ganeshkhind, Pune 411 007, India}

{\footnotesize nkd@iucaa.ernet.in}

\vskip 1.3truecm

\textsl{Josep M. Pons}

\textit{DECM and ICC, Facultat de F\'{\i}sica, Universitat de Barcelona,\\ Diagonal 647, E-08028 Barcelona,
Catalonia, Spain.}

\textit{\&}

\textsl{IUCAA, Post Bag 4, Ganeshkhind, Pune 411 007, India}

{\footnotesize pons@ecm.ub.es}

\vskip 2truecm

\textbf{Abstract}
\end{center}

\bigskip

{\small In the framework of the Einstein-Palatini formalism, even though the projective transformation connecting the arbitrary connection with the Levi Civita connection has been floating in the literature for a long time and perhaps the result was implicitly known in the affine gravity community, yet as far as we know Julia and Silva were the first to realise its gauge character. We rederive this result by using the Rosenfeld-Dirac-Bergmann approach to constrained Hamiltonian systems and do a comprehensive self contained analysis establishing the equivalence of the Einstein-Palatini and the metric formulations without having to impose the gauge choice that the connection is symmetric. We also make contact with the the Einstein-Cartan theory when the matter Lagrangian has fermions.}

\vskip 1truecm

\vfill\eject

%%%%%%%%%%%%%%%%%%%%%%%%%%%%%%%%%%%%%%%%%%%%%%%
\section{Introduction}
\label{intro}
%%%%%%%%%%%%%%%%%%%%%%%%%%%%%%%%%%%%%%%%%%%%%%%

Consider pure general relativity (GR) with the connection as an additional independent field. There are two usual formulations available, namely the Einstein-Palatini (EP) or affine-metric formulation (see \cite{Ferraris:1982} for a historical account) and the Vielbein-Einstein-Palatini (VEP) formulation in which the independent variables are the vielbein and the spin-connection. In both cases it is well known, see for instance, \cite{trau,Sandberg:1975db,hehl,stachel,Floreanini:1990kt,Percacci:1990wy} that the solution for the connection differs from the Levi Civita connection by a projective transformation which is a symmetry of the GR Lagrangian in both EP and VEP formulations. With the aim to regain the standard GR, these authors as well as other practitioners in the field modify the original setting in several ways so as always to end up with the connection dynamically becoming the Levi Civita connection. The usual procedure is to restrict the original connection in order to get rid of the projective transformation. This  is the standard textbook approach which is also used in \cite{trau}. In this case one considers torsion free connection in the EP formalism or metric compatible connection (Einstein-Cartan's theory) in the VEP formalism. Other authors choose to modify the original Lagrangian by adding a term with a Lagrange multiplier that eventually forces the connection to become the Levi Civita \cite{hehl,stachel}. Another procedure is to add to the Lagrangian new terms quadratic in the torsion and the non-metricity tensor in such a way that the Levi Civita connection comes about dynamically \cite{Floreanini:1990kt,Percacci:1990wy}.

\vspace{4mm}

In fact, there is no need to indulge into any such manipulations because the projective transformation is indeed a gauge transformation. As far as we know, this result was first given in \cite{Julia:1998ys} in the case of pure gravity. We think we can add to this result some new contributions, which will define the contents of our present paper: {\bf a)} we provide with a new constructive derivation of the gauge symmetry, now obtained by using the Rosenfeld-Dirac-Bergmann (RDB) approach to gauge theories (constrained systems); {\bf b)} we show that in the case of pure gravity the presence of this gauge symmetry imposes that the only connection that can be present in an observable must be the Levi Civita connection; {\bf c)} we extend the results of \cite{Julia:1998ys} to the case with fermions and show that, from the Einstein-Palatini point of view, the Einstein-Cartan theory is the natural extension of GR with fermionic matter; and {\bf d)} we prove that the usual procedure beginning with a torsion free connection in the EP formalism, or a connection with metricity in the VEP formalism, is a consistent truncation of the original setting and it amounts to eliminating the gauge symmetry of the projective transformation but does not reduce the physical degrees of freedom.

\vspace{4mm}

It would be fair to say that the projective transformation has been floating
in the works of many authors (see the review \cite{Hehl:1994ue}) yet its gauge
character has largely remained unnoticed particularly in the GR community
despite Ref.\cite{Julia:1998ys}. The main focus of this paper is to bring out this aspect
with emphasis and clarity, and then to carry out the constraint analysis which
is certainly new and insightful.

The relevance of examining the consequences of the gauge character of the projective transformations relies on the fact that the physical obervables in a theory endowed with gauge symmetry must be gauge invariant. In our case - even including matter Lagrangians not depending on the connection - we will show that the only connection that can participate in an observable is the Levi Civita connection. If there is fermionic matter with the standard minimal coupling to gravity, we show that the connection that can participate in an observable is metric compatible though it might have torsion.

\vspace{4mm}

Ever since the original formulation by Einstein \cite{ein25} of EP formalism, it was in fact known that it was sufficient to have the vanishing trace of the torsion tensor for reducing the arbitrary connection to the Levi Civita connection; i.e to be free of the projective transformation. In our view this condition is nothing but a gauge fixing condition whereas the stronger condition of the vanishing of the torsion, still in the EP formalism, is a consistent truncation of the original theory with free connection which also eliminates the projective transformation. We will find the analogue of Einstein's gauge fixing for the VEP formalism and will also show that metric compatibility is in this case a consistent truncation with the same effects as in the EP formalism. The message is that the EP and VEP formulations with arbitrary and free connection are fully equivalent to GR. The equivalence also holds for matter Lagrangians not involving the connection. In the case of fermions with minimal coupling we show that the VEP formalism is equivalent to the Einstein-Cartan theory.

\vspace{4mm}

We intend to make the paper comprehensive and self contained. We start in the next Section by rederiving Einstein's result in a very simple manner for the EP case. Next we turn to the VEP case in Section 3 where we find the precise requirement on the spin-connection that plays a similar role for the VEP formulation. In Section 4, we explain the origin of these conditions as a result of the gauge fixing of a $R^d$ gauge symmetry - the projective transformation - first discovered in \cite{Julia:1998ys}, though the result might have implicitly been known in the affine geometry community. The  clear message that emerges is that the Einstein GR equation would always follow regardless of whether one chooses the gauge condition or not in both EP \cite{Sotiriou:2009xt} and VEP formulations. We also show in this section that in the pure gravity case the only connection that can participate in an observable is the Levi Civita connection. In Section 5 we give an independent construction of the projective gauge symmetry by carrying out the canonical analysis within the framework of the RDB approach to gauge theories and identifying the associated first class constraints that generate the projective transformations. The counting of degrees of freedom is also discussed in this section. In Section 6, we consider the inclusion of fermions where we find metricity with torsion (Einstein-Cartan theory). Finally we conclude with some remarks.

%%%%%%%%%%%%%%%%%%%%%%%%%%%%%%%%%%%%%%%%%%%%%%%
\section{The EP formulation with an arbitrary connection}
\label{EP}\setcounter{equation}{0}
%%%%%%%%%%%%%%%%%%%%%%%%%%%%%%%%%%%%%%%%%%%%%%%

Here we consider the EP Lagrangian
\beq
\bar{\cal L}_{\!{}_{E\!P}} =\sqrt{-g}g^{\mu\nu} \bar R_{\mu\nu}
\eeq
where
\beq
\bar R_{\mu\nu}= \partial_\sigma \bar \Gamma_{\ \mu\nu}^\sigma
-\partial_\mu \bar \Gamma^\sigma_{\ \sigma\nu}
+ \bar \Gamma^\rho_{\ \mu\nu}\bar \Gamma^\sigma_{\ \sigma\rho}-\bar \Gamma^\rho_{\ \sigma\nu}\bar \Gamma^\sigma_{\ \mu\rho}\,
\label{rmunu}
\eeq
is the Ricci tensor for $\bar \Gamma$, an arbitrary affine connection. The overhead bar is indicative of the connection being an arbitrary independent field and the quantities built from it. Our aim is to determine $\bar \Gamma$ by means of its own EOM which is, as it stands, a differential equation. One can easily turn the problem into an algebraic one by the simple change of variables
\beq
\bar \Gamma_{\  \mu\nu}^\sigma = \Gamma_{\ \mu\nu}^\sigma + C_{\ \mu\nu}^\sigma\,
\label{changvar}
\eeq
where $\Gamma$ is the Levi-Civita connection (its components are the Christoffel symbols) determined unambiguously from the metric under the conditions of metricity and torsionfree. The Lagrangian becomes a functional of the metric and the tensor $C_{\ \mu\nu}^\sigma$,
\beq
\bar{\cal L}_{\!{}_{E\!P}}= \sqrt{-g}g^{\mu\nu}(R_{\mu\nu} -\nabla_\mu C_{\ \rho\nu}^\rho
+\nabla_\rho C_{\ \mu\nu}^\rho + C_{\ \mu\nu}^\lambda C_{\ \rho\lambda}^\rho -C_{\ \rho\nu}^\lambda C_{\ \mu\lambda}^\rho)\,,
\eeq
with the covariant derivative relative to the Levi-Civita connection which is also used to build the Ricci tensor $R_{\mu\nu}$. The terms with the covariant derivative are just divergences which will not affect the Euler-Lagrange EOM. Since we are only interested in the EOM, we drop these terms in the following and we end up with the Lagrangian
\beq
\bar{\cal L}_{\!{}_{E\!P}}={\cal L}_{\!{}_{E\!H}} + \sqrt{-g}g^{\mu\nu}(C_{\ \mu\nu}^\lambda C_{\ \rho\lambda}^\rho -C_{\ \rho\nu}^\lambda C_{\ \mu\lambda}^\rho) + {\rm divergence}\,,
\label{potential}
\eeq
where ${\cal L}_{\!{}_{E\!H}}$ stands for the Einstein-Hilbert Lagrangian. Thus the EP Lagrangian becomes, up to a divergence, the EH Lagrangian plus a potential term which does not depend on the spacetime derivatives. The EOM for $C$ is
\beq
g^{\mu\nu}\frac{\partial }{\partial C_{\ \alpha\beta}^\gamma} (C_{\ \mu\nu}^\lambda C_{\ \rho\lambda}^\rho -C_{\ \
\rho\nu}^\lambda C_{\ \mu\lambda}^\rho) = g^{ \alpha\beta}C_{\ \rho\gamma}^\rho +\delta_\gamma^\alpha
C^{\beta\rho}_{\ \ \,\rho} - C^{\alpha\beta}_{\ \ \, \gamma}- C^{\beta\ \, \alpha}_{\ \, \gamma}=0\,.
\eeq
So we must solve the algebraic equation
\beq
g_{ \alpha\beta}C_{\ \rho\gamma}^\rho + g_{\alpha\gamma} C_{\beta\rho}^{\ \ \,\rho}- C_{\alpha\beta \gamma}-
C_{\beta\gamma\alpha}=0\,.
\label{algeq}
\eeq
With some algebraic manipulation, as shown in the appendix, we obtain
\beq
C^{\alpha}_{\ \beta \gamma}=\delta^{\alpha}_{\gamma}U_\beta\,,
\label{cfinal}
\eeq
for an arbitrary vector $U_\beta$. It is clear that for the connection to dynamically become the Levi Civita connection, what is required is $U_\mu=0$. Further by using eqs (\ref{changvar}) and (\ref{cfinal}), the torsion tensor can be written as
$$
\bar T_{\ \mu\nu}^\sigma = T_{\ \mu\nu}^\sigma = C_{\ \mu\nu}^\sigma-C_{\ \nu\mu}^\sigma = \delta^\sigma_\nu U_\mu -
\delta^\sigma_\mu U_\nu\,,
$$
and its trace gives $U_\mu =\frac{1}{d-1}\bar  T_{\ \mu\nu}^\nu$, which should vanish to yield the usual EOM (i.e., determining the connection as the Levi Civita connection). This is the condition Einstein obtained in 1925 \cite{ein25},
\beq
 U_\mu = \bar T_{\ \mu\nu}^\nu=0.
\label{EPcond}
\eeq
and pronounced that it is necessary to assume this condition to get to the usual EOM. It should be noted that what is required is the vanishing of the trace of the torsion tensor and not of the torsion itself.

%%%%%%%%%%%%%%%%%%%%%%%%%%%%%%%%%%%%%%%%%%%%%%%
\section{The VEP formulation with an arbitrary spin-connection}
\label{VEP}\setcounter{equation}{0}
%%%%%%%%%%%%%%%%%%%%%%%%%%%%%%%%%%%%%%%%%%%%%%%

In this case we consider the Lagrangian as
\beq
\bar{\cal L}_{\!{}_{V\!E\!P}} = \vert e \vert e^{[\mu}_I e^{\nu] J} \bar R_{\mu\nu  \ J}^{\ \ \ I},
\label{VEP-lag}
\eeq
where $e_I = e^\mu_I \partial_\mu$ is the vielbein ($e^\mu_I e^\nu_J \eta^{IJ} = g^{\mu\nu}\,,\ e_\mu^I =
g_{\mu\nu}e^\nu_J \eta^{IJ} =(e^\nu_J)^{-1}$) and
$$
\bar R_{\mu\nu  \ J}^{\ \ \ I}= \partial_\mu \bar \omega_{\nu \  J}^{\ \, I} + \bar \omega_{\mu \  K}^{\ \,
I}\,\bar\omega_{\nu \  J}^{\ \, K} - (\mu \leftrightarrow \nu)\,
$$
is the Riemann tensor expressed as a $2$-form. The spin-connection $1$-form $\bar\omega_{\mu \  J}^{\ \, I}$ is a functional of the vielbein and the
connection $\bar \Gamma_{\  \mu\nu}^\sigma$, defined by using the property that the covariant derivative of the identity tensor, $I = e^I_\nu \, dx^\nu \otimes e^{}_{I}\,,$ must vanish (it is sometimes referred to as the ``tetrad postulate''),
\beq
\bar\nabla_\mu e^I_\nu = \partial_\mu e^I_\nu  - \bar\Gamma_{\ \mu\nu}^\sigma e^I_\sigma + \bar\omega_{\mu \  J}^{\
\, I} e^J_\nu=0.
\label{covderid}
\eeq
Now we take the vielbein $e^\mu_I$ and the spin-connection $\bar\omega_{\mu \  J}^{\ \, I}$ as independent variables.
No conditions are imposed on the spin-connection. We proceed in the same way as in the previous Section and define
$D_{\mu \  J}^{\ \, I}$ by
\beq
\bar\omega_{\mu \  J}^{\ \, I}=\omega_{\mu \  J}^{\ \, I}+D_{\mu \  J}^{\ \, I}\,
\label{changvar2}
\eeq
where $\omega_{\mu \  J}^{\ \, I}$ is the Levi Civita spin-connection, i.e, torsionfree and metric compatible.

Note that with an arbitrary spin-connection, we are not implementing local Lorentz gauge transformations, but instead implementing local $GL(d,R)$ gauge transformations. Infinitesimal transformations are defined as $\delta M^I = a^I_{\ J}M^J$ and
$\delta N_I = - a^J_{\ I}N_J$ so that $\delta (M^I N_I) =0$. Note also that $\delta \eta^{IJ}=a^{IJ}+ a^{JI}$, which vanishes only for the Lorentz transformations. Thus if we want to keep the Minkowski metric invariant we must restrict to the local Lorentz group. This is what we will do henceforth.

\vspace{4mm}

Proceeding similarly to the previous Section, the Lagrangian becomes
\beq
\bar {\cal L}_{\!{}_{V\!E\!P}} = {\cal L}_{\!{}_{V\!E\!H}} + \vert e \vert e^{[\mu}_I e^{\nu] J}(D_{\mu \  K}^{\ \, I}D_{\nu \  J}^{\ \, K}-D_{\nu \  K}^{\ \, I}D_{\mu \  J}^{\ \, K})+ {\rm divergence}\,,
\label{potential2}
\eeq
and to study the EOM for $D$ we need only to consider the piece
\beq
e^{\mu}_I e^{\nu J}(D_{\mu \  K}^{\ \, I}D_{\nu \  J}^{\ \, K}-D_{\nu \  K}^{\ \, I}D_{\mu \  J}^{\ \, K})
= E^{\mu\sigma}_{\ \ \,\sigma}E^{\rho}_{\ \rho\mu}-E^{\mu}_{\ \rho\sigma}E^{\rho\sigma}_{\ \ \,\mu}\,,
\label{piece}
\eeq
where we have carefully defined $D_{\rho \  K}^{\ \, I}=: e^\mu_K e^I_\nu E_{\mu\rho}^{\ \ \,\nu}$ in order to match eq. (\ref{piece}) with the potential term in eq. (\ref{potential}), with $E\leftrightarrow C$. In view of eq. (\ref{cfinal}) the solution for $E$ is therefore $E^{\alpha}_{\ \beta \gamma}=\delta^{\alpha}_{\gamma}V_\beta$ with $V_\beta$ arbitrary. Going back to the $D$ variables, it takes the analogous form
\beq
D_{\mu \  J}^{\ \, I} = \delta^I_J V_\mu\,.
\label{dfinal}
\eeq

Thus we are again led to the same gauge condition $V_\mu=0$ for the spin-connection to dynamically become the Levi Civita connection. This condition is equivalent to  $D_{\mu \  I}^{\ \,  I}=0$, which is equivalent to
\beq
\bar\omega_{\mu \  I}^{\ \, I}=0\,
\label{VEPcond}
\eeq
which is analogous to the Einstein condition (\ref{EPcond}). This would then dynamically imply metricity and torsionfree character for the connection in the VEP formulation. The invariance of $\bar\omega_{\mu \  I}^{\ \, I}$ under the local Lorentz transformation is obvious because the flat indices are saturated. On the other hand,  $\bar\omega_{\mu \  I}^{\ \, I}$ is a $1$-form under diffeomorphism and therefore the equation (\ref{VEPcond}) is geometric.

%%%%%%%%%%%%%%%%%%%%%%%%%%%%%%%%%%%%%%%%%%%%%%%
\section{The $R^d$ gauge symmetry of projective transformations}
\label{u1}\setcounter{equation}{0}
%%%%%%%%%%%%%%%%%%%%%%%%%%%%%%%%%%%%%%%%%%%%%%%

%%%%%%%%%%%%%%%%%%%%%%%%%%%%%%%%%%%%%%%%%%%%%%%
\subsection{The gauge group of projective transformations and its gauge fixing}
\label{newgaugegroup}\setcounter{equation}{0}
%%%%%%%%%%%%%%%%%%%%%%%%%%%%%%%%%%%%%%%%%%%%%%%

Apart from the diffeomorphism invariance and the local Lorentz invariance in the VEP case,
eqs (\ref{cfinal}) and (\ref{dfinal}) indicate to the existence of new gauge symmetry in the EP and
VEP formalisms. Let us start by considering first the EP formalism. Since $U_\beta$ in eq.\! (\ref{cfinal}) is arbitrary, this means that
$C_{\ \mu\nu}^\sigma\rightarrow C_{\ \mu\nu}^\sigma + \delta^\sigma_{\nu} U_\mu$
must be a symmetry of the EOM. In fact it is a Noether gauge symmetry that leaves the
Lagrangian $\bar{\cal L}_{\!{}_{E\!P}}$ invariant, as one can easily verify. This is the
symmetry of projective transformations already mentioned in Section \ref{intro}.
Although scarcely noticed in the previous literature, it is clearly a gauge symmetry because the
parameters of the symmetry are arbitrary functions of the spacetime coordinates\footnote{We will further argue this important point in Section \ref{canon}}. In turn this means that $\bar{\cal L}_{\!{}_{E\!P}}$ is invariant under the $R^d$ gauge group defined by the finite transformation
\beq
 g_{\mu\nu} \rightarrow g_{\mu\nu}\,,\  \quad
\bar \Gamma_{\  \mu\nu}^\sigma \rightarrow \bar\Gamma_{\  \mu\nu}^\sigma+\delta^\sigma_{\nu} U_\mu\,,
\label{gaugeEP}
\eeq
for an arbitrary vector field $U_\mu$ \cite{Sotiriou:2009xt}. As we said in Section \ref{intro} this transformation (\ref{gaugeEP}) has indeed been recognized by many authors as a
symmetry of the Lagrangian but surprisingly its gauge character has however remained overlooked, with the  notable exception of \cite{Julia:1998ys}. All other authors drew the erroneous inference that the GR EOM could only be obtained by imposing the constraint (\ref{EPcond}) or by the other procedures mentioned in Section \ref{intro}. This is however not true and it would be explicitly shown in the next Section.

Now we can give a new meaning to the equation (\ref{EPcond}): {\it it is just a good gauge fixing condition for the $R^d$ gauge symmetry. This is something which had not been realized in any of the previous discussions of this problem.}

Note that introducing from the scratch a symmetric connection, that is, torsionless, appears now as a consistent truncation of the theory\footnote{ We take the strong sense in which a truncation is said to be consistent if the truncated EOM of the original Lagrangian coincide with the EOM of the truncated Lagrangian.} and its only effect is of eliminating the gauge symmetry of the projective transformations while retaining the physical degrees of freedom unaffected.

\vspace{4mm}

We now turn to the VEP case. Keeping eq. (\ref{dfinal}) in mind, we see that the $R^d$ gauge symmetry
is expressed through $D_{\mu I J} \rightarrow D_{\mu I J}+ \eta_{I J} V_\mu$, which leaves $\bar{\cal
L}_{\!{}_{\!V\!E\!P}}$ invariant. Thus the Lagrangian ${\cal L}_{\!{}_{\!V\!E\!P}}$ is invariant under
the finite gauge transformation
\beq
e^{\mu}_I \rightarrow e^{\mu}_I\,,\  \quad
 \bar\omega_{\mu \  J}^{\ \, I}\rightarrow \bar\omega_{\mu \  J}^{\ \, I}+\delta^I_J V_\mu\,.
\label{gaugeVEP}
\eeq
In this perspective, eq.\! (\ref{VEPcond}) is a good gauge fixing condition for the VEP formalism.
Eqs (\ref{gaugeEP}) and (\ref{gaugeVEP}) are in fact the same as they refer to the same gauge symmetry. Indeed, eq. (\ref{covderid}) implies the well known formula
$
\bar\omega_{\mu \  J}^{\ \, I} =e^I_\nu(\partial_\mu e^\nu_J + \bar \Gamma_{\
\mu\rho}^\nu e^\rho_J)\,,
$
from which it follows eq. (\ref{gaugeEP}) $\Leftrightarrow$ eq. (\ref{gaugeVEP}), with $V_\mu=U_\mu$. It is worth noting though that the gauge conditions (\ref{EPcond}) and (\ref{VEPcond}) are different but each of them is holonomous in its own variables.

Similarly in the EP case, note that imposition of the antisymmetry of the spin-connection, i.e. making it metric compatible from the scratch, is a consistent truncation of the theory which just eliminates the gauge symmetry of the projective transformation. The physical degrees of freedom remain the same.

%%%%%%%%%%%%%%%%%%%%%%%%%%%%%%%%%%%%%%%%%%%%%%%
\subsection{The equation of motion without gauge fixing}
\label{eomnogauge}
%%%%%%%%%%%%%%%%%%%%%%%%%%%%%%%%%%%%%%%%%%%%%%

The next question arises: what happens when no gauge is fixed? The answer is that we still obtain the standard GR. Let us work in the EP case with the original variables $g$ and $\bar\Gamma$. Note that
\beq
\frac{\delta\bar{\cal L}_{\!{}_{E\!P}}}{\delta C} = 0 \quad  \Leftrightarrow  \quad\frac{\delta\bar{\cal L}_{\!{}_{E\!P}}}{\delta
\bar\Gamma}{\vert}_{\bar\Gamma= \Gamma + C} = 0.
\eeq
It follows from eq. (\ref{algeq}) that the connection EOM is solved to give
\beq
\frac{\delta\bar{\cal L}_{\!{}_{E\!P}}}{\delta \bar\Gamma} = 0 \quad\Leftrightarrow \quad
\bar\Gamma^{\alpha}_{\ \beta \gamma}=\Gamma^{\alpha}_{\ \beta \gamma}+\delta^{\alpha}_{\gamma}U_\beta\,,
\label{eom2}
\eeq
with $U_\beta$ arbitrary.

\vspace{4mm}

On the other hand, the metric EOM for ${\cal L}_{\!{}_{E\!P}}$, including matter only coupled to the metric, is
\beq
\bar R_{(\mu\nu)} - \frac{1}{2} g_{\mu\nu} \bar R = T_{\mu\nu}\,,
\label{eom1}
\eeq
where only the symmetric part of the Ricci tensor is relevant. Under the substitution dictated by the solution to eq. (\ref{eom2}), the Ricci tensor becomes
$$ \bar R_{\mu\nu}{\vert{}_{(\ref{eom2})}} = R_{\mu\nu} + 2\partial_{[\mu}U_{\nu]}\,.
$$
Putting this into eq. (\ref{eom1}), we end up with
\beq
R_{\mu\nu} - \frac{1}{2} g_{\mu\nu} R = T_{\mu\nu}\,.
\label{eom3}
\eeq
That is the standard EOM for GR. This result was already noticed in \cite{Sotiriou:2009xt,gia}. Let us mention that the considerations of the non-observability of the projective transformation also emerge in the work of \cite{ehlerschild,schouten,Fatibene:2011fy}.

We conclude that the EOM for ${\cal L}_{\!{}_{E\!P}}$ can be written as eq. (\ref{eom3}) which is just GR equation together with the equation for the connection eq. (\ref{eom2}). In here the $R^d$ gauge freedom is fully manifest and its physical meaning we now analyze. To construct physical observables\footnote{ For a discussion on observables in generally covariant theories, see \cite{Rovelli:1990ph,Pons:2009cz}.} we must resort to gauge invariant quantities. Comparing eqs (\ref{gaugeEP}) and (\ref{eom2}), we infer that any gauge invariant quantity built up by using the connection must be independent of the gauge parmeters $U_\mu$. This means that the same quantity could be built by using just the Levi Civita connection which is indeed invariant under the $R^d$ gauge group instead of the original one. {\sl This shows that it is the Levi Civita connection which is physically meaningful}.

\vspace{4mm}

On a historical note, let us mention that Einstein was convinced that the only way to obtain GR was by requiring the connection to dynamically become the Levi Civita connection in the first place. He made this claim very clear in \cite{ein25} (see the English translation) by saying:
{\it Had we not assumed the vanishing of the $\phi_{\tau}$
{\rm ($\phi_{\tau}$ is our $U_{\tau}$)}, we would have been unable to derive the known law
of the gravitational field in the above manner by assuming the symmetry of the $g_{\mu \nu}$.}
With the benefit of hindsight, we may say that he (as well as other authors, among them  \cite{hehl,stachel}) did not notice the $R^d$ gauge symmetry, else he would have seen that the vanishing of $\phi_\tau$ is in fact a gauge condition,  our eq. (\ref{EPcond}), which did not have to be assumed to obtain eq. (\ref{eom3}).

\vspace{4mm}

Similarly the same analysis could be carried through for the VEP formalism. In fact eqs
(\ref{changvar2}) and (\ref{dfinal}) imply $\bar\omega_{\mu \  J}^{\ \, I}=\omega_{\mu \  J}^{\ \,
I}+\delta^I_J V_{\mu}$\, which in turn implies $\bar R_{\mu\nu  \ J}^{\ \ \, I}=R_{\mu\nu  \ J}^{\ \ \,
I} +2 \delta^I_J \partial_{[\mu}V_{\nu]}$. In view of the antisymmetry of the flat indices in the Lagrangian (\ref{VEP-lag}), it is clear that the
Lagrangian is invariant under the $R^d$ gauge group and so is the vielbein EOM. Therefore the vielbein EOM coincides with the standard GR EOM in the vielbein formalism.

\vspace{4mm}

A different line of thought was developed in \cite{Floreanini:1990kt,Percacci:1990wy} in which the VEP Lagrangian
is extended with terms quadratic in the torsion and the non-metricity tensor. This
extension is natural from an effective field theory standpoint. When the change of variables
(\ref{changvar2}) is applied, the Lagrangian becomes that of EH plus terms quadratic in the $D$
variables. The generic form of these terms is of the type $Q^{\mu I J\nu K L} D_{\mu I  J}D_{\nu K  L}$. For the non-degenerate matrix $Q$ it turns out that EOM for $D$ would require vanishing of these variables which means the theory is dynamically equivalent to the standard EH. It suggests that inclusion of the quadratic terms seems to be effectively equivalent to fixing the gauge in our case.

\vspace{4mm}

The analysis above and in the previousl subsection shows that the EP and the VEP formalisms, with the usual restrictions of vanishing torsion and metric compatibility respectively, are completely equivalent to Einstein's GR even including matter as long as the matter Lagrangian does not depend on the connection. This is because the connection becomes dynamically the Levi Civita connection in both the cases. If there are fermions, the matter Lagrangian depends on the connection which then asks for the vielbein formulation and in that case the resulting VEP formalism does not yield the Levi Civita connection. However the observable connection is still metric compatible. More on this later.

%%%%%%%%%%%%%%%%%%%%%%%%%%%%%%%%%%%%%%%%%%%%%%%
\section{Canonical analysis and degrees of freedom}
\label{canon}\setcounter{equation}{0}
%%%%%%%%%%%%%%%%%%%%%%%%%%%%%%%%%%%%%%%%%%%%%%%
In this section we will use the techniques of the theory of constrained systems as
developed by Rosenfeld, Dirac and Bergmann,
\cite{rosenfeld30,bergmann49a,bergbrun49,bergm3,dirac50,dirac4} (References of books include
\cite{Sundermeyer:1982gv,Henneaux:1992ig,Gitman:1990qh}. See \cite{Pons:2004pp} for a
brief introduction to the RDB theory). This Section can be read as a completely independent derivation of the projective transformations symmetry and its gauge character.

\vspace{4mm}

The change of variables (\ref{changvar}) offers another nice advantage that it is easy to do the
canonical phase space analysis of the theory. Dropping the boundary term in eq. (\ref{potential}),
irrelevant for our purpose, the Lagrangian under consideration is
\beq \bar{\cal L}_{\!{}_{E\!P}}={\cal L}_{\!{}_{E\!H}} + \sqrt{-g}g^{\mu\nu}(C_{\ \mu\nu}^\lambda C_{\
\rho\lambda}^\rho -C_{\ \rho\nu}^\lambda C_{\ \mu\lambda}^\rho)\,.
\label{potential3}
\eeq
The canonical Hamiltonian is just the $ADM$ Hamiltonian \cite{Arnowitt:1962hi} with a new potential term,
\beq
\bar{\cal H}_{\!{}_{E\!P}} = {\cal H}_{\!{}_{A\!D\!M}} - \sqrt{-g}g^{\mu\nu}(C_{\ \mu\nu}^\lambda C_{\
\rho\lambda}^\rho -C_{\ \rho\nu}^\lambda C_{\ \mu\lambda}^\rho)\,,
\eeq
in which we should replace $g^{\mu\nu}$ components according to the $ADM$ decomposition, with lapse $N$ and shift $N^{j}$,
\beq g_{\mu\nu}= \left(\begin{array}{c|c}- N^2+N^i \gamma_{ij} N^{j}  & \gamma_{ij} N^{j} \\ \hline \gamma_{ji} N^{i} & \gamma_{ij}
\end{array}\right)\,.
\eeq
The primary constraints of the theory are $d\ $ momenta $P_\mu\simeq 0$ canonically conjugate to the lapse and shift together with
$d^3\ $ momenta $\Pi^{\ \mu\nu}_\lambda \simeq 0$ canonically conjugate to the variables $C_{\ \mu\nu}^\lambda$. We use Dirac notation of weak equality, $\simeq$, to indicate that the constraints vanish when the EOM, now in phase space, is satisfied.
The dynamics in phase space is given by the Dirac Hamiltonian, which is the canonical Hamiltonian plus additional terms linear in the primary constraints with Lagrange multipliers,
$$\bar{\cal H}_{\!{}_D}=\bar{\cal H}_{\!{}_{E\!P}} + \lambda^\mu P_\mu + \lambda_{\ \mu\nu}^\rho\Pi^{\ \mu\nu}_\rho\,.
$$
Let us first look for the secondary constraints arising from the requirement that the constraints $\Pi^{\ \mu\nu}_\rho$ are preserved in time. The equations are
$$ \{\Pi^{\ \mu\nu}_\rho,\, \bar{\cal H}_{\!D}\} \simeq 0\,
$$
which give equation (\ref{algeq}) as the new secondary constraints. This is however equivalent to eq. (\ref{cfinal}) and it describes the independent constraints
\beq
C_{\ \mu\nu'}^\nu \simeq 0\ ({\rm with}\ \nu'\neq \nu)\,,\quad C_{\ \mu 0}^0 - C_{\ \mu i}^i\simeq 0\,,\ i= 1,\cdots d-1\,,
\label{seccons}
\eeq
which are $d^2(d-1) + d(d-1) = d(d^2 - 1)\ $ secondary constraints. In addition there are another $d\ $ secondary constraints implied by the preservation in time of the constraints $P_\mu$. These are the well known $ADM$ Hamiltonian and momentum constraints, now with additional pieces coming from the potential term in eq. (\ref{potential3}). It is worth noting that these additional pieces can be safely eliminated because
the potential term, $\sqrt{-g}g^{\mu\nu}(C_{\ \mu\nu}^\lambda C_{\ \rho\lambda}^\rho -C_{\ \rho\nu}^\lambda C_{\ \mu\lambda}^\rho)$ is proportional to the constraints (\ref{seccons}). No tertiary constraints arise in the formalism.

Consider the constraints $\Pi^{\ \mu\nu}_\lambda$ together with the constraints (\ref{seccons}). A simple observation tells that they organize together as $2d(d^2 - 1)\ $ second class constraints (defining a locally simplectic submanifold) and a set of $d\ $ remaining first class constraints. These first class constraints are given by the trace
$\Pi^{\ \mu\nu}_\nu$ which have vanishing Poisson bracket with the constraints (\ref{seccons}).
What we have found are just the generators of the $R^d$ gauge algebra. In fact, defining the gauge generator
$$
G = \int d^{d-1}x\, \epsilon_\mu \Pi^{\ \mu\nu}_\nu\,
$$
for an infinitesimal set of arbitrary\footnote{ Note that this arbitrariness allows for functions of compact support. In the case of diffeomorphism invariance, this compact support was instrumental for Einstein's hole argument. See \cite{Fatibene:2010cf} for the relation of the hole argument with gauge transformations.} functions $\epsilon_\mu$, we obtain
$$
\delta C^{\alpha}_{\ \beta \gamma} := \{ C^{\alpha}_{\ \beta \gamma} ,\,G\}= \delta^{\alpha}_{\gamma} \epsilon_\beta\,
$$
which is the infinitesimal gauge transformation of the $R^d$ gauge group described in eq. (\ref{gaugeEP}) and the discussion above it. Notice that this result can be taken as a new independent derivation of the projective transformation as symmetry of the theory  including the proof of its gauge character because it is generated by first class constraints.

As regards the diffeomorphism symmetry (general convariance) one can find the gauge generators following on the lines of
\cite{Pons:1996av}. This is not a trivial task because the formalism now must account for the action of the diffeomorphism on the $C$ variables in phase space. We will not pursue this task here.

Let us count the degrees of freedom. In addition to the standard GR counting, we have  $2 d^3\ $ new variables in phase space, the $C$'s and the $\Pi$'s, but they are constrained by $2d(d^2 - 1)\ $ second class constraints plus $d\ $ first class constraints, the latter generating the $R^d$ gauge transformation. To gauge fix the transformation we need $d\ $ gauge conditions like eq. (\ref{EPcond}). Thus the total number of constraints plus the gauge condition exactly match the number of new variables introduced, $2 d^3\ $. With all this squared out we are again left with the GR degrees of freedom.
%%%%%%%%%%%%%%%%%%%%%%%%%%%%%%%%%%%%%%%%%%%%%%%
\section{Including fermions}
\label{ferm}\setcounter{equation}{0}
%%%%%%%%%%%%%%%%%%%%%%%%%%%%%%%%%%%%%%%%%%%%%%%

Now we will include a matter Lagrangian with the standard Dirac fermion\footnote{Of course, one can add to the picture a Maxwell field with minimal coupling with the fermion but this addition does not involve the spin-connection and hence will have no effect.}. This would however be true for any other spinor with the same type of coupling to gravity. The Lagrangian is
$$\bar{\cal L}_{\!{}_M} = \vert e \vert ({\rm i\,}e_I^\mu \bar \psi \gamma^I  \bar\nabla_\mu \psi - m \bar \psi\psi )\,,
$$
with
$$
\bar\nabla_\mu \psi^a = \partial_\mu \psi^a + \frac{1}{2}\bar \omega_{\mu K L}(\Sigma^{K L})^a_b \psi^b\,,
$$
where $\Sigma^{KL}$ is the $(\frac{1}{2},0) + (0,\frac{1}{2})$ representation of the Lorentz algebra. Notice that the spin-connection does not need to be antisymmetric in the flat indices but only its antisymmetric component couples to the fermion regardless of the specific representation it is in.

The local Lorentz gauge group is realized for the total Lagrangian
${\cal L}_{\!{}_{V\!E\!P}}+ {\cal L}_{\!{}_M}$ with the standard definitions for the transformation of the spinors and the spin-connection under the Lorentz group.

We proceed along the lines of Section \ref{VEP} and define $D_{\mu I  J}$ by
\beq
\bar\omega_{\mu I  J} = \omega_{\mu I  J}+D_{\mu I  J}\,
\label{substD}
\eeq
with $\omega_{\mu I  J}$ being the Levi Civita spin-connection. Note that $D_{\mu I  J}$ appears linearly in ${\cal L}_{\!{}_M}$ as
$ D_{\mu I  J} A^{\mu I  J}\,$ with
$$
A^{\mu I  J}:= \frac{\rm i}{2}\vert e\vert e_K^\mu\bar \psi   \gamma^K \Sigma^{I J}  \psi \,.
$$
The $D$-dependent part in the total Lagrangian ${\cal L}_{\!{}_{V\!E\!P}}+ {\cal L}_{\!{}_M}$ is
$$e^{\mu}_I e^{\nu J}(D_{\mu \  K}^{\ \, I}D_{\nu \  J}^{\ \, K}-D_{\nu \  K}^{\ \, I}D_{\mu \  J}^{\ \, K})+D_{\mu I  J} A^{\mu I  J}\,.
$$
Defining $\ E_{\mu\rho}^{\ \ \,\nu}:= D_{\rho \  K}^{\ \, I}e_\mu^K e_I^\nu\ $ and
$\ A^{\mu\rho\sigma}:= e^\rho_K e^\sigma_L A^{\mu K L}$ we can write the above expression as
$$E^{\mu\sigma}_{\ \ \,\sigma}E^{\rho}_{\ \rho\mu}-E^{\mu}_{\ \rho\sigma}E^{\rho\sigma}_{\ \ \,\mu} + E^\sigma_{\ \mu\rho} A^{\mu\rho}_{\ \ \,\sigma}\,
$$
with $A^{\mu\rho\sigma}$ being antisymmetric in the last two indices. The EOM for the $E$ variables is
\beq
g_{ \alpha\beta}E_{\ \rho\gamma}^\rho + g_{\alpha\gamma} E_{\beta\rho}^{\ \ \,\rho}- E_{\alpha\beta \gamma}-
E_{\beta\gamma\alpha}+ A_{\alpha\beta \gamma}=0\,.
\label{algeq2}
\eeq
Following on the same lines as in the appendix for the case with no fermions, the solution to the above equation is found to be
\beq
E_{\alpha\beta \gamma} = \frac{1}{2} (A_{\alpha\beta \gamma}-A_{\beta \gamma\alpha}-A_{\gamma\beta\alpha }) +
\frac{1}{d-2}(g_{\alpha\beta} A_{\rho\gamma}^{\ \ \,\rho}-g_{\gamma\beta} A_{\rho\alpha}^{\ \ \,\rho}) + g_{\alpha\gamma}U_\beta\,
\label{solfer}
\eeq
where $U_\beta$ is arbitrary. In terms of the $D$ variables, the result is
\beq
D_{\beta I J} = -\frac{1}{2} A_{\beta I J} +  A_{\mu K L}\Big(\frac{1}{2}e^K_\beta(e^\mu_J \delta^L_I-e^\mu_I \delta^L_J) + \frac{1}{d-2}e^{\mu L}(e_{J\beta} \delta^K_I-e_{I\beta} \delta^K_J)\Big)+\eta_{I J}U_\beta\,.
\label{solferflat}
\eeq
Note that it decomposes into an antisymmetric and a symmetric part, the latter being just a gauge artifact.

Having in mind eq. (\ref{substD}) we infer from eq. (\ref{solferflat}) the following:
\begin{itemize}
 \item
The gauge group $R^d$ is still present in the formalism\footnote{The existence of the projective transformation is also recognized in \cite{Hehl:1994ue} but with no mention of its gauge character.}.
\item
Up to a gauge transformation of $R^d$, the spin-connection is antisymmetric in its flat indices.
\item
The gauge condition (\ref{VEPcond}) is still valid, when implemented it eliminates the $R^d$ gauge freedom and yields metricity for the spin-connection.
\item
Putting eq. (\ref{solferflat}) into eq. (\ref{substD}) and next plugging it into the vielbein EOM for $\bar{\cal L}_{\!{}_{V\!E\!P}}+ \bar{\cal L}_{\!{}_M}$, the final equation is the same as if we had started with an antisymmetric spin-connection as an input from the outset. This is because the symmetric part of the spin-connection, $\eta_{I J}U_\beta$, which is merely a gauge artifact, does not enter into the veilbein EOM.
\item
We observe that metricity, equivalently the antisymmetry of the spin-connection, is carried through for the fermionic matter while the torsionfree condition is violated by the antisymmetric component in the right hand side of eq. (\ref{solferflat}). This is the  Einstein-Cartan theory of gravity in which fermions source torsion (see \cite{Hammond:2002rm} for a general account). In this theory metricity is a  precondition whereas in here it arises dynamically from the arbitrary free connection.
\end{itemize}

%%%%%%%%%%%%%%%%%%%%%%%%%%%%%%%%%%%%%%%%%%%%%%%
\section{Conclusion}
\label{conc}\setcounter{equation}{0}
%%%%%%%%%%%%%%%%%%%%%%%%%%%%%%%%%%%%%%%%%%%%%%%

In most of the previous discussions of the Einstein-Palatini formulation (in particular \cite{trau,Sandberg:1975db,hehl,stachel,Floreanini:1990kt,Percacci:1990wy,Sotiriou:2009xt}), it is clear that the gauge character of the transformation (\ref{gaugeEP}) has not been realized. In fact it was first stated in \cite{Julia:1998ys}. In this paper we construct this gauge symmetry by using canonical methods, and then use it to gain a clearer understanding of the physical equivalence between the Einstein-Hilbert and the Einstein-Palatini formulations of general relativity. Besides let us also note in the following some points which are interesting and insightful.

\vspace{4mm}

We have shown that the Einstein condition (\ref{EPcond}) is a gauge condition of the $R^d$ gauge symmetry admitted by the EOM. This is indeed insightful that when we come to the Palatini formulation, there is an increase in the number of degrees of freedom which have to be controlled by the connection EOM. This is what precisely appears as the gauge condition  which is simply to make the torsion tracefree. Of course the gauge condition should have no effect on the metric equation of motion and that is what is done in Ref.\! \cite{Sotiriou:2009xt} and also shown in Section 4.2. However Einstein had not realized this symmetry nor its gauge character and that is why he had to assume it as a condition to get the GR EOM. In fact he believed that the GR EOM could be retrieved only if the connection became dynamically the Levi Civita connection. On the other hand Trautman \cite{trau} does recognise the arbitrariness in determination of the connection signified by the projective transformation but overlooks spotting of its gauge character.

Further we do the canonical analysis in phase space and work out its full structure of constraints. It is then shown that the additional degrees of freedom introduced by the EP formalism are exactly countered by the constraints and the new $R^d$ gauge symmetry with its gauge condition. We have also written down the generator of this symmetry.

\vspace{4mm}

We have also done the entire analysis in the VEP case and shown that there is a duality between the basic variables and their corresponding relations for the EP and VEP formalisms. This duality has been already discussed in \cite{Percacci:1990wy}. In the EP case the torsionfree condition is algebraic, a symmetry property of the connection, whereas metricity is a differential relation of the type, $\  \partial g + \Gamma g+ \Gamma g=0$. While for the VEP, it is the metricity that is algebraic, an antisymmetry property of the spin-connection, whereas the torsionfree condition is a differential relation of the type, $\  \partial e + \omega e=0$. Now we extend this duality to the gauge fixing conditions: being holonomous in their respective settings, the gauge conditions (\ref{EPcond}) and (\ref{VEPcond}) are consistent with this duality \cite{Percacci:1990wy}. Still in the VEP formalism, it is shown that the fermions could also be included in the matter Lagrangian which would however require non-zero torsion but metricity is preserved. This is Einstein-Cartan theory of gravity. Of course there is nothing new in saying that fermions are sources for torsion. The main point here is that metricity is an {\sl outcome} of the dynamics instead of being postulated form the outset as is usually done. Our result thus makes the case for the claim that Einstein-Cartan theory is the natural extension of Einstein's GR when fermionic matter is present.

\vspace{4mm}

We have seen that, now without fermions, the connection EOM dynamically determine it as the Levi-Civita connection up to a gauge transformation which does not affect the metric EOM. Alternatively one can plug back the solution of the connection EOM into the Lagrangian itself which would give the standard EH action having the symmetric Levi-Civita connection. We should however point out here that it is this connection that is physically observable in the parallel transport, and the reason is the following. To construct physical observables we must resort to gauge invariant quantities. Noticing from (\ref{eom2}) that $\bar\Gamma_{\  \mu\nu}^\sigma = \Gamma_{\ \mu\nu}^\sigma + \delta^\sigma_\nu U_\mu$, we infer that any gauge invariant quantity built up by using the connection\footnote{When $\bar\Gamma_{\  \mu\nu}^\sigma$ is plugged into the geodesic equation there would arise an additional term, $(U_\mu dx^\mu/ds)dx^\sigma/ds$ which could always be absorbed by reparametrization of the affine parameter \cite{wald}. Thus the geodesic equation would remain covariant under this transformation.} must be independent of the gauge parmeters $U_\mu$. This means that the same quantity may be built by using just the Levi Civita connection instead of the original one. This shows that it is the Levi Civita connection which is physically meaningful. This argument is trivially extended to the fermionic case, where the metric compatible connection - now with torsion - is the physically meaningful connection.

\vspace{4mm}

We have also shown that the usual procedures which are followed in textbooks and the standard discussion of this problem of beginning with a symmetric connection, i.e. torsion free in the EP formalism, or a connection with metricity in the VEP formalism, are both consistent truncations of the general case of arbitrary free connection and it simply amounts to eliminating the gauge symmetry of the projective transformation without reducing  the physical degrees of freedom.

\vspace{4mm}

Finally from a conceptual and intuitive standpoint, we find it extremely satisfying and appealing that the EP-VEP formulations with matter not coupled to the connection (while in the VEP fermions could be included but they would generate torsion with metricity retained) are entirely equivalent to the EH formulation with no a priori condition on the connection. We would however like to emphasize the new realization that in the EP-VEP formulation, new degrees of freedom get introduced which are then controlled by the gauge symmetry of projective transformations and the corresponding gauge fixing condition is exactly what Einstein had to assume to get the standard gravitational equation. The gauge fixing condition is however never a necessary assumption. Paraphrasing in the MTW idiom \cite{MTW}, the connection indeed {\it flaps the breeze} full fledged, even more so than what the MTW asked for, and dynamically collapses to the standard gravitational dynamics.

%%%%%%%%%%%%%%%%%%%%%%%%%%%%%%%%%%%%%%%%%%%%%%%
\section*{Appendix}
\label{app}\setcounter{equation}{0}
%%%%%%%%%%%%%%%%%%%%%%%%%%%%%%%%%%%%%%%%%%%%%%%

Here we solve equation (\ref{algeq}) in detail.
The tensor $C$ has three traces, which we label $c_1\,,c_2\,,c_3\,,$ according to the position of the remaining free
index
$$
c_1 \leftrightarrow C_{\alpha\rho}^{\ \ \,\rho}\,,\ c_2 \leftrightarrow C^{\rho}_{\ \,\alpha \rho}\,,\
c_3 \leftrightarrow C^{\rho}_{\ \, \rho\alpha}\,.
$$
Then, taking traces on (\ref{algeq}) we obtain ($d$ is the spacetime dimension ($d>2$)) the two relationships
$$
(d-1) c_3 -c_2 + c_1 =0 \,, \qquad  (d-1) c_1 -c_2 + c_3 =0\,,
$$
from which $c_3 = c_1 $ follows. Defining the vector $C_\alpha :=C_{\alpha\rho}^{\ \ \,\rho}= C^{\rho}_{\ \,
\rho\alpha}$, one has, for (\ref{algeq}),
\beq
C_{\alpha\beta \gamma}+ C_{\beta\gamma\alpha}= g_{\alpha\beta}C_\gamma + g_{\alpha\gamma}C_\beta\,.
\label{ccycl}
\eeq
Writing the cyclic permutations, $\alpha\beta \gamma\rightarrow\gamma\alpha\beta\rightarrow\beta\gamma\alpha$, we
obtain a total of three relations, the first one being (\ref{ccycl}). Subtracting the second relation from the first
one and adding up the third, we obtain as a final result $C_{\alpha\beta \gamma}= g_{\alpha\gamma}U_\beta$, for an arbitrary vector $U_\beta$ (note that taking traces on this last result, we recover $U_\beta=C_\beta$.). Raising the first index we obtain (\ref{algeq}).

%%%%%%%%%%%%%%%%%%%%%%%%%%%%%%%%%%%%
\section*{Acknowledgments}
%%%%%%%%%%%%%%%%%%%%%%%%%%%%%%%%%%%%
We would like to thank T. Padmanabhan for pointing out to us Ref.\! \cite{Sotiriou:2009xt} and for discussions, and Lu\'{\i}s Navarro for help regarding the references to Einstein's papers. We also thank Romesh Kaul for an insightfull discussion and an anonymous referee for pointing out the earlier Ref.\! \cite{trau} on the projective transformation, however its gauge character remained unspotted. We also thank anonymous referees for very constructive criticism that has considerably improved the presentation of the results and in particular bringing to our notice the important  Ref.\! \cite{Julia:1998ys}. JMP acknowledges support by MCYT FPA 2007-66665, CIRIT GC 2005SGR-00564 and Spanish Consolider-Ingenio 2010 Programme CPAN (CSD2007-00042). He also acknowledges IUCAA and its people for the warm hospitality.


\begin{thebibliography}{99}

%\cite{Ferraris:1982}
\bibitem{Ferraris:1982}
  M.~Ferraris, M.~Francaviglia and C.~Reina,
``Variational formulation of general relativity from 1915 to 1925 ``Palatini's method'' discovered by Einstein in 1925,''
  Gen.\ Rel.\ Grav.\  {\bf 14}, 243 (1982)

\bibitem{trau}
A.~Trautman, ``On the structure of the Einstein-Cartan equations``.  Symposia Mathematica, Vol. XII (Convegno di Relativit\`a, INDAM, Rome, 1972),  pp. $139 - 162$. Academic Press, London, 1973.

%\cite{Sandberg:1975db}
\bibitem{Sandberg:1975db}
  V.~D.~Sandberg,
 ``Are torsion theories of gravitation equivalent to metric theories?,''
  Phys.\ Rev.\  D {\bf 12} (1975) 3013.
  %%CITATION = PHRVA,D12,3013;%%


%\cite{hehl}
\bibitem{hehl}
F.~W.~Hehl and G.~D.~Kerlick,
``Metric-Affine Variational Principles in General
Relativity. I. Riemannian Space-Time,''
Gen.\ Rel.\ Grav. {\bf 9} (1978) 691

\bibitem{stachel}
A.~Papapetrou and J.~Stachel,
``A New Lagrangian for the Vacuum Einstein Equations and Its Tetrad Form,''
Gen.\ Rel.\ Grav. {\bf 9} (1978) 1075


%\cite{Floreanini:1990kt}
\bibitem{Floreanini:1990kt}
  R.~Floreanini and R.~Percacci,
  ``Palatini formalism and new canonical variables for GL(4) invariant
  gravity,''
  Class.\ Quant.\ Grav.\  {\bf 7} (1990) 1805.
  %%CITATION = CQGRD,7,1805;%%

%\cite{Percacci:1990wy}
\bibitem{Percacci:1990wy}
  R.~Percacci,
  ``The Higgs Phenomenon in Quantum Gravity,''
  Nucl.\ Phys.\  B {\bf 353} (1991) 271
  [arXiv:0712.3545 [hep-th]].
  %%CITATION = NUPHA,B353,271;%%

%\cite{Julia:1998ys}
\bibitem{Julia:1998ys}
  B.~Julia and S.~Silva,
  ``Currents and superpotentials in classical gauge invariant theories. 1.
  Local results with applications to perfect fluids and general relativity,''
  Class.\ Quant.\ Grav.\  {\bf 15} (1998) 2173
  [arXiv:gr-qc/9804029].


%\cite{Hehl:1994ue}
\bibitem{Hehl:1994ue}
  F.~W.~Hehl, J.~D.~McCrea, E.~W.~Mielke and Y.~Ne'eman,
  ``Metric affine gauge theory of gravity: Field equations, Noether identities,
  world spinors, and breaking of dilation invariance,''
  Phys.\ Rept.\  {\bf 258} (1995) 1
  [arXiv:gr-qc/9402012].
  %%CITATION = PRPLC,258,1;%%

\bibitem{ein25}
A.~Einstein, ``Einheitliche Fieldtheorie von Gravitation und Elektrizit\"at``, Pruess. Akad. Wiss. {\sl 414}, 1925; A.~Unzicker and T.~Case,
``Translation of Einstein's attempt of a unified field theory with teleparallelism,''arXiv:physics/0503046.

%\cite{Sotiriou:2009xt}
\bibitem{Sotiriou:2009xt}
T.~P.~Sotiriou,
``f(R) gravity, torsion and non-metricity,''
Class.\ Quant.\ Grav.\  {\bf 26} (2009) 152001
[arXiv:0904.2774 [gr-qc]]. We thank T. Padmanabhan for pointing out this reference to us. In fact we had independently found the same result before knowing of it.

%\cite{gia}
\bibitem{gia}
  G.~Giachetta and L.~Mangiarotti,
  ``Projective invariance and Einstein Equations,''
  arXiv:1010.0869 [gr-qc].



%\cite{ehlerschild}
\bibitem{ehlerschild}
J.~Ehlers, F.~A.~E.~Pirani and A.~Schild,
``The geometry of free fall and light propagation,'' in {\sl General Relativity}, ed. L. Oraifeartaigh (Claredon, Oxford, 1972)

%\cite{schouten}
\bibitem{schouten}
J.A.Schouten,
``Ricci calculus: an introduction to Tensor Analysis and its Geometrical Applications.''
Springer Verlag 1954

%\cite{Fatibene:2011fy}
\bibitem{Fatibene:2011fy}
  L.~Fatibene, M.~Francaviglia and G.~Magnano,
  ``On a Characterization of Geodesic Trajectories and Gravitational Motions,''
  arXiv:1106.2221 [gr-qc].
  %%CITATION = ARXIV:1106.2221;%%

%\cite{Rovelli:1990ph}
\bibitem{Rovelli:1990ph}
  C.~Rovelli,
  ``What is observable in classical and quantum gravity?,''
  Class.\ Quant.\ Grav.\  {\bf 8} (1991) 297.
  %%CITATION = CQGRD,8,297;%%

%\cite{Pons:2009cz}
\bibitem{Pons:2009cz}
  J.~M.~Pons, D.~C.~Salisbury and K.~A.~Sundermeyer,
  ``Revisiting observables in generally covariant theories in the light of
  gauge fixing methods,''
  Phys.\ Rev.\  D {\bf 80}, 084015 (2009)
  [arXiv:0905.4564 [gr-qc]].
  %%CITATION = PHRVA,D80,084015;%%


%\bibitem{Pauli}
%W.~Pauli,
%Theory of Relativity, Pergamon Press. 1958. Translated from the article %"Relati\-vi\-t\"atstheorie", Supplementary notes, Encyklop\"adie der mathematischen %Wissenschaften, Vol. V19, (B. G. Teubner, Leipzig 1921)

%\bibitem{EK}
%A. ~Einstein and B. ~Kaufman, ``A new form of the general relativistic field equations,''
%Ann.\ Math., Princeton, {\bf 62} (1955) 128
\bibitem{rosenfeld30}
L.~Rosenfeld, ``Zur Quantelung der Wellenfelder''  Annalen der Physik {\bf  397}, 113-152, (1930).

\bibitem{bergmann49a} P. G. Bergmann,
``Non-Linear Field Theories,''  Phys.\ Rev.\  {\bf 75} (1949), 680 - 685.

\bibitem{bergbrun49}
P. G. Bergmann and J. H. M. Brunings,
`` Non-Linear Field Theories II. Canonical Equations and Quantization,''
 Rev.\ Mod.\ Phys. {\bf 21} (1949) 480 - 487.

\bibitem{bergm3}
J.~L.~Anderson and P.~G.~Bergmann,
``Constraints In Covariant Field Theories,''
Phys.\ Rev.\  {\bf 83} (1951) 1018.

\bibitem{dirac50}
P.A. M. Dirac, ``Generalized Hamiltonian Dynamics,''  Can.\ J.\ Math.\ {\bf 2}, (1950) 129 - 148

\bibitem{dirac4} P. A. M. Dirac,
 ``Lectures on Quantum Mechanics,''
    Yeshiva Univ.\ Press, New York (1964).

%\cite{Sundermeyer:1982gv}
\bibitem{Sundermeyer:1982gv}
  K.~Sundermeyer,
  ``Constrained Dynamics With Applications To Yang-Mills Theory, General
  Relativity, Classical Spin, Dual String Model,''
  Lect.\ Notes Phys.\  {\bf 169} (1982) 1.

%\cite{Henneaux:1992ig}
\bibitem{Henneaux:1992ig}
  M.~Henneaux and C.~Teitelboim,
  ``Quantization of gauge systems,''
 Princeton, USA: Univ. Pr. (1992) 520 p

%\cite{Gitman:1990qh}
\bibitem{Gitman:1990qh}
  D.~M.~Gitman and I.~V.~Tyutin,
  ``Quantization of fields with constraints,''
%\href{http://www.slac.stanford.edu/spires/find/hep/www?irn=2399601}{SPIRES entry}
{\it  Berlin, Germany: Springer (1990) 291 p. (Springer series in nuclear and particle physics)}

%\cite{Pons:2004pp}
\bibitem{Pons:2004pp}
  J.~M.~Pons,
  ``On Dirac's incomplete analysis of gauge transformations,''
  Stud.\ Hist.\ Philos.\ Mod.\ Phys.\  {\bf 36} (2005) 491
  [arXiv:physics/0409076].


%\cite{Arnowitt:1962hi}
\bibitem{Arnowitt:1962hi}
  R.~L.~Arnowitt, S.~Deser and C.~W.~Misner,
  ``The dynamics of general relativity,'' in
{\sl Gravitation: an introduction to current research}, Louis Witten ed. (Wilew 1962), 1962.
  arXiv:gr-qc/0405109.


%\cite{Pons:1996av}
\bibitem{Pons:1996av}
  J.~M.~Pons, D.~C.~Salisbury and L.~C.~Shepley,
  ``Gauge transformations in the Lagrangian and Hamiltonian formalisms of
  generally covariant theories,''
  Phys.\ Rev.\  D {\bf 55} (1997) 658.
[arXiv:gr-qc/9612037].
  %%CITATION = PHRVA,D55,658;%%

%\cite{Fatibene:2010cf}
\bibitem{Fatibene:2010cf}
  L.~Fatibene, M.~Francaviglia and S.~Mercadante,
  ``Noether Symmetries and Covariant Conservation Laws in Classical,
  Relativistic and Quantum Physics,''
  arXiv:1001.2886 [gr-qc].
  %%CITATION = ARXIV:1001.2886;%%


%\cite{Hammond:2002rm}
\bibitem{Hammond:2002rm}
  R.~T.~Hammond,
  ``Torsion Gravity,''
  Rept.\ Prog.\ Phys.\  {\bf 65} (2002) 599.
  %%CITATION = RPPHA,65,599;%%


\bibitem{MTW}
C.~W.~Misner, K.~S.~Thorne and J.~A.~Wheeler,
 ``Gravitation,'' Freeman and Co., San Francisco,  1973

\bibitem{wald}

R.~M.~Wald, ``General Relativity,'' Overseas Press (India) Pvt. Ltd., 2007.


\end{thebibliography}
\end{document}